\documentclass[aps,prb,twocolumn,superscriptaddress]{revtex4}
\usepackage{amssymb}
\usepackage{amsmath}
 
\usepackage[pdftex]{graphicx}

\usepackage{amsfonts}
\usepackage{bbold}
\usepackage{color}

\begin{document}

\title{Tunneling into and between helical edge states - fermionic approach}
\author{D.N. Aristov}
\affiliation{NRC ``Kurchatov Institute", Petersburg Nuclear 
Physics Institute, Gatchina 188300, Russia}
\affiliation{Department of Physics, St.Petersburg State University, Ulianovskaya 1,
St.Petersburg 198504, Russia}
\author{R.A. Niyazov}
\affiliation{NRC ``Kurchatov Institute", Petersburg Nuclear 
Physics Institute, Gatchina 188300, Russia}
\affiliation{Department of Physics, St.Petersburg State University, Ulianovskaya 1,
St.Petersburg 198504, Russia}
\date{\today}

\begin{abstract}
We study four-terminal junction of spinless Luttinger liquid wires, which describes either a corner junction of two helical edges states of topological insulators or the tunneling from the spinful wire into the helical edge state. We use the fermionic representation and the scattering state formalism, in order to compute the renormalization group (RG) equations for the linear response conductances. 
We establish our approach by considering a junction between two possibly non-equivalent helical edge states and find an agreement with the earlier analysis of this situation.  Tunneling from the tip of the spinful wire to the edge state is further analyzed which requires some modification of our formalism. 
In the latter case we demonstrate i) the existence of both fixed lines and conventional fixed points of RG equations, and ii) certain proportionality relations holding for conductances during renormalization. The scaling exponents and phase portraits are obtained in all cases. 
\end{abstract}

\pacs{71.10.Pm, 73.63.-b, 71.10.Hf}
\maketitle

\section{Introduction}
    

The advances in technology stimulate a renewed theoretical interest to the properties of one dimensional (1D) quantum wires.  The practical implementations of such systems include carbon nanotubes,
chains of metal atoms or weakly side-coupled molecular chains in solids. A new class of materials is given by two-dimensional topological insulators (TIs), whose edge states are ideal quantum wires ~\cite{Hasan2010, Qi2011}.  This paper discusses the transport via the junction between these edge states and its renormalization by interactions.

The transport properties of 1D wires has been theoretically studied since early 1990s   in terms of Luttinger liquid and within bosonization formalism \cite{Giamarchi1988,Kane1992,Furusaki1993}. It was found that the interaction between fermions renormalizes the impurity scattering, so that for repulsive interaction the conductivity of the wire tends to zero in the limit of low temperatures, even for one impurity. The bosonization approach considers the interaction in the bulk of the wire exactly, while the impurity is regarded as perturbation. In more sophisticated theories the impurity is considered as certain boundary conditions for the bulk fields, and the approach is generalized for the case of junctions between several wires \cite{Oshikawa2006}. 

Alternative   description using more traditional fermionic approach and S-matrix formalism was developed in \cite{Matveev1993a,Yue1994} for case of one impurity,  this approach was later generalized to the case of junctions of several wires \cite{Lal2002,Das2004}.  It was found that the renormalization of S-matrix by interaction is described in terms of renormalization group equations, which eventually defined the scaling exponents of the conductance behavior.  The initial formulation of the fermionic S-matrix approach assumed that the bulk interaction was considered only in the lowest order, and this might be regarded as some disadvantage in comparison with bosonization. In certain cases the lowest order calculation was insufficient, and the calculation of next-order corrections to S-matrix was required \cite{Teo2009,Aristov2010}

It was found, however,  that the S-matrix approach could be essentially improved by summation of  simple diagrammatic series  in perturbation theory \cite{Aristov2009,Aristov2013}. The result of this summation was the modification of the RG equations for S-matrix, so that the scaling exponents found in limiting fixed points (FPs) of these equations coincided with those established by bosonization.  At the same time the S-matrix approach has an advantage over bosonizaton in providing full scaling curves for the conductances.  
The method was tested for junctions of  two leads \cite{Aristov2009}, three-lead
junctions \cite{Aristov2010,Aristov2011a,Aristov2013}, also for non-equilibrium \cite{Aristov2014}. Generalization of this approach for the case of infinite Luttinger liquid wires was undertaken  in \cite{Shi2016}. 

It was recently shown that the case of a junction connecting four quantum wires, which is generally characterized by S-matrix belonging to $U(4)$ group, might be principally different in the form of RG equation, even in the lowest order of interaction. \cite{Aristov2015b} This difference from the simpler cases of junctions of two and three wires shows itself in the discrete $Z_{2}$ ambiguity in the parametrization of S-matrix, which is unobservable in terms of conductances of the junction. This ``hidden phase'' is known to happen in $U(N)$ groups at $N\ge 4$, and it may indicate that the bosonization description becomes inadequate already for four wires. 

In this  paper we continue our studies of junctions of four quantum wires. First we analyze   
the renormalization of corner junction  between the 1D helical edge states of topological insulators, earlier discussed in  \cite{Hou2009,Teo2009,Stroem2009,Huang2013}.  The work by Teo and Kane \cite{Teo2009} allows the direct comparison with our analysis here, as they provide the second-order RG analysis for S-matrix,  in addition to the bosonization treatment.  It was argued in   \cite{Teo2009} that the time-reversal symmetry determines the specific structure of the S-matrix, in particular the absence of backscattering in the same edge state. We extend the analysis of Ref.\  \cite{Teo2009}  by summing the perturbation series in bulk interaction and arrive to non-perturbative RG equations for conductances, whose FPs and scaling exponents are in exact correspondence with bosonization results, where available. The above problem of ``hidden'' phase in $U(4)$ is absent in case of helical edge states due to the symmetry of S-matrix, reducible to the symplectic form. 
  
Next we analyze the generalization of the above setup by considering different interaction strength in the helical edge states, whereas the form of the S-matrix remains the same due to symmetry. The phase portrait in this case is characterized by two phases; the new phase as compared to the previous case appears for interaction strength of different signs, when some FPs disappear or change their character. 
We show that if the X-junction is characterized by the ``bare'' S-matrix close to one of FPs of saddle point type, then the temperature evolution reveals the non-monotonic behavior of conductances, similarly to the case of Y-junction considered in \cite{Aristov2010}.  

We further notice that the case of electron tunneling from a spinful Luttinger liquid wire tip to helical edge state can be described by the same S-matrix, but different matrix of bulk interactions. This physically important setup was not considered previously and it is different from the tunneling from fully spin-polarized tip to the helical edge state  \cite{Das2011}, the latter situation described in terms of three-lead Y-junction.  
The different form of the interaction matrix leads to slightly modified derivation of RG equation, but otherwise our approach remains the same.   As a result, we obtain the phase portrait describing the tunneling from the spinful wire to the helical edge state. We find only one fixed point, corresponding to the absence of tunneling. In addition, we demonstrate one or two  {fixed lines} of conductances, depending on the interaction. 
The RG fixed lines in the plane of  conductances was found earlier for chiral Y-junctions at certain values of the interaction strength. \cite{Aristov2012a}  The scaling exponents are in exact agreement with those expected from bosonization arguments. 
 
The rest of the paper is organized as follows. 
We describe our method  in the Section \ref{sec:model}. The setup of our model, the particular form of S-matrix and the relevant set of conductances are introduced, and the renormalization equations are briefly discussed here.  In Section \ref{sec:flows}  we discuss tunneling between helical edge states, first in a simpler case of symmetric junction, and then introducing asymmetry in interaction strength.  In Section \ref{sub:probe}  we discuss tunneling from a spinful wire to the edge state. This case requires some modification of our method and we sketch the corresponding derivation. We present the concluding remarks in Section \ref{sec:conclu}.
 
\section{The model and renormalization of conductances
\label{sec:model}}

\subsection{The model}
We consider the  two-channel Tomonaga-Luttinger liquid (TLL) model with a local scatterer of arbitrary strength 
 in the middle of the wire. In particular, such model incorporates a model of spinful electrons, scattering on a local potential and a model of corner junction between the helical edge states of topological insulators.  

As in our previous papers, we assume that the short-range interaction between the fermions is of the forward scattering type and takes place in wires of finite length $L$, contacted by reservoirs. The adiabatic
transition from wire to reservoir  produces
no additional potential scattering. The junction is assumed to have
microscopic extension $l$ of the order of the Fermi wave length.  Inside the
junction interaction effects are neglected. This is expressed below by the window
function $\Theta(x)=1$, if $l<x<L$, and zero otherwise. The regions $x>L$ are thus regarded as reservoirs or leads labeled $j=1,2,3,4$. 

For the linearized spectrum near the Fermi energy we may write the TLL Hamiltonian in the representation of incoming and outgoing waves in channel $j$ (fermion operators $\psi _{j,in}$, $\psi _{j,out}$) as

\begin{equation}
\begin{aligned}
\mathcal{H} &=\int_0^{\infty}dx[H^{0}+H^{int} \Theta( \ell < x < L )]\,,   \\
H^{0} &= v_{F} \Psi_{in}^{\dagger }i\nabla \Psi _{in}-v_{F}\Psi
_{out}^{\dagger }i\nabla \Psi _{out}\,, \\
H^{int} &= 2\pi v_{F}  \sum\limits_{j,k=1}^4  g_{jk}
 \widehat{\rho }_{j} \widehat{\widetilde{\rho }}_{k} \,.  
\end{aligned} 
\label{Ham}
\end{equation}

Here $\Psi _{in}=(\psi _{1,in},\psi _{2,in},\psi _{3,in},\psi _{4,in})$ denotes a vector
operator of incoming fermions and the corresponding vector of outgoing
fermions is expressed through the $S$-matrix as $\Psi _{out}(x)=S\cdot \Psi
_{in}(x)$ at $x\to 0$.  
We put the Fermi velocity $v_{F}=1$ below. 
The interaction term of the Hamiltonian is expressed in
terms of density operators $\widehat{\rho }_{j,in}=\Psi ^{+}\rho _{j}\Psi =%
\widehat{\rho }_{j}$, and $\widehat{\rho }_{j,out}=\Psi ^{+}\widetilde{\rho }%
_{j}\Psi =\widehat{\widetilde{\rho }}_{j}$, where $\widetilde{\rho }%
_{j}=S^{+}\cdot \rho _{j}\cdot S$ and the density matrices are given by $%
(\rho _{j})_{\alpha \beta }=\delta _{\alpha \beta }\delta _{\alpha j}$ and $(%
\widetilde{\rho }_{j})_{\alpha \beta }=S_{\alpha j}^{+}S_{j\beta }$. 
The 4$\times$4 unitary $S$-matrix characterizes the scattering at the junction and belongs to $U(4)$ group.  
Depending on the physics of the problem, the form of the  $S$-matrix and the interaction matrix, $g_{jk}$, varies.

\begin{figure} 
\includegraphics[width=0.48\columnwidth]{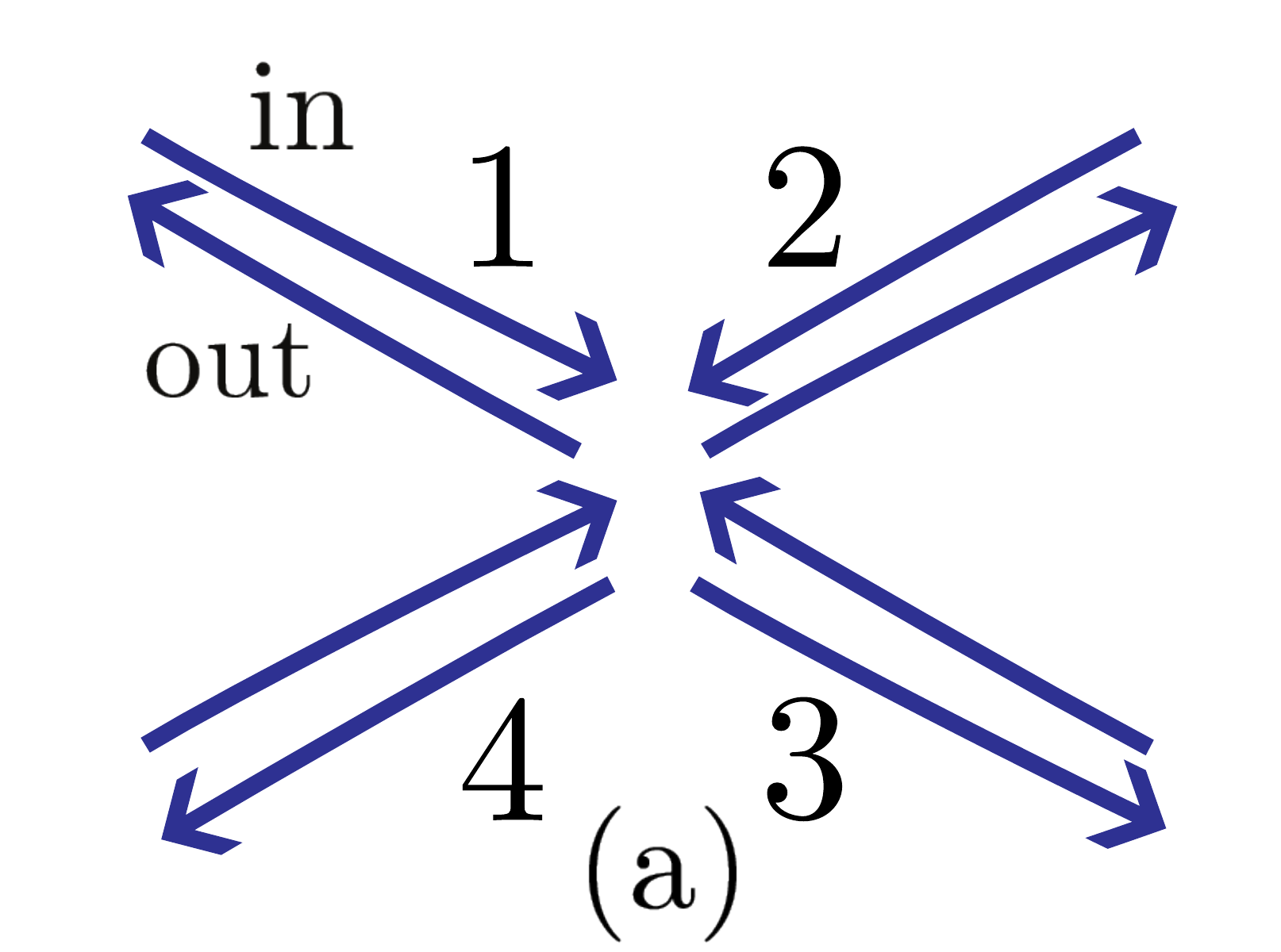} \includegraphics[width=0.48\columnwidth]{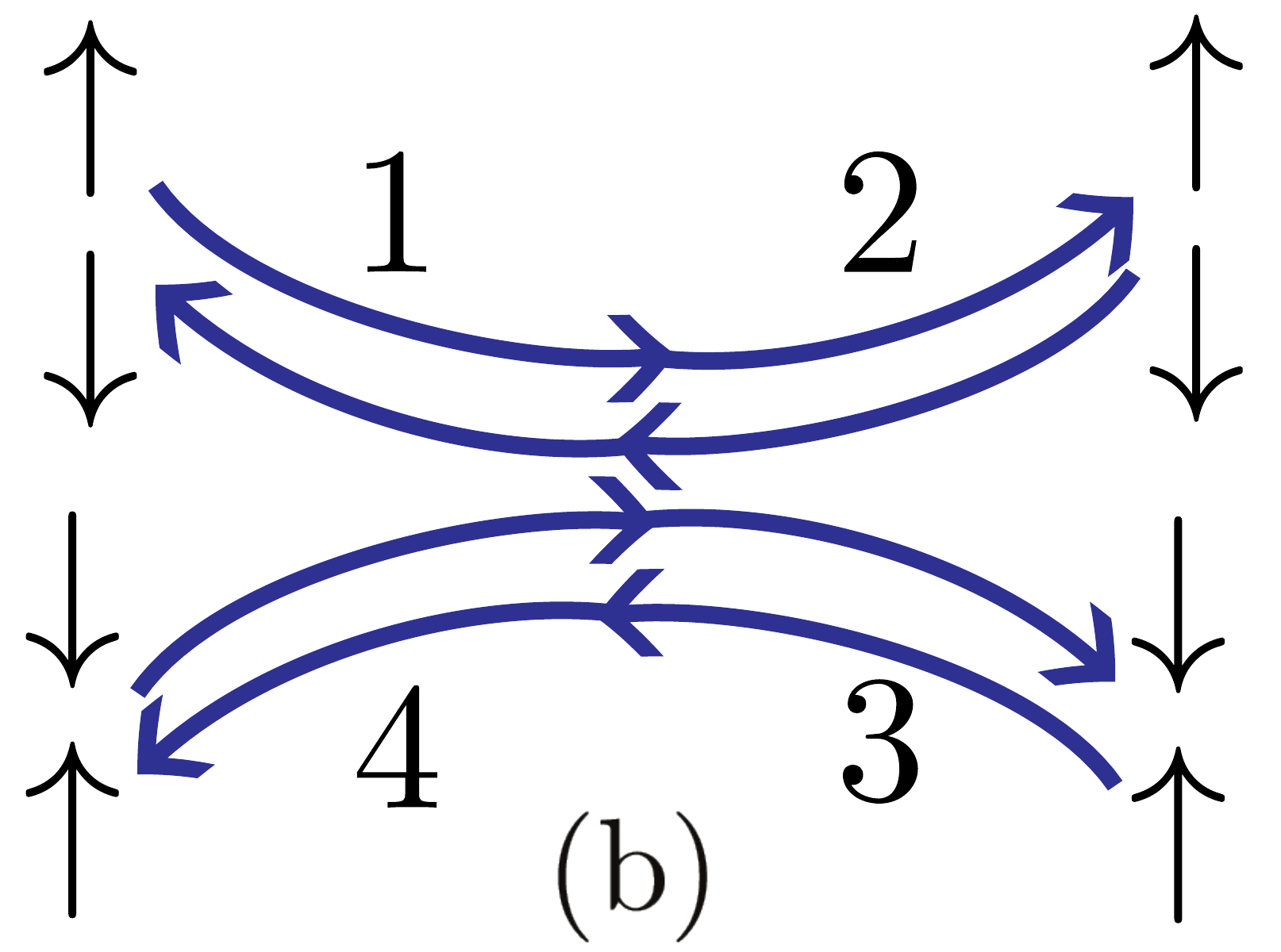} 
\caption{\label{fig:spinful}
(a) Schematically shown X-junction, (b) X-junction of the helical edge states, with spin projections explicitly indicate, see also  Table \ref{tab:cor2}.  }
\end{figure}

Specifically, we consider below the corner junction between the helical edge states in symmetric and asymmetric setup with respect to interaction. We also consider the  tunneling of electrons from the spinful wire tip into the helical edge state.   
A general geometry of X-junction is shown in Fig. \ref{fig:spinful}a.  It turns out, that all the above cases with spinful fermions can be described by the Fig.\  \ref{fig:spinful}b, and the difference between these cases is encoded in the form of $g_{jk}$.  For further convenience, we explicitly give the correspondence between the channel and the spin index  in Table \ref{tab:cor2}.

\begin{table}
\caption{\label{tab:cor2}Correspondence between the helical edge states with projections of spin and the numbering of our channels.}
\begin{ruledtabular}
\begin{tabular}{l|lllll}
& 1& 2& 3& 4  \\
\hline
in &   $\uparrow$,  left &  $\downarrow$,  right  & 
  $\uparrow$,  right & $\downarrow$,   left  \\
out & $\downarrow$, left & $\uparrow$,   right & 
$\downarrow$,    right & $\uparrow$,   left \\  
\end{tabular}
\end{ruledtabular}
\end{table}

\subsection{Reduced conductances}

In the linear response regime our system is characterized by the matrix of conductances defined by $I_i=C_{ij} V_j$, with  the current $I_i$ flowing in channel $i$ and the  voltage $V_j$ applied to the channel $j$.  The current conservation, $\sum I_i=0$, and the absence of response to the equal change in voltages lead to  
the Kirchhof's rules, $\sum _{i}C_{ij} = \sum _{j}C_{ij} = 0$. 
It means that we can choose more convenient linear combinations of  $I_i$, $V_{j}$ reducing the number of independent components in $C_{ij}$.  Using the Kubo formula, one has in the d.c.\ limit  $C_{ij}=\tfrac12 (\delta_{ij}-Y_{ij}) $, with $Y_{ij} = |S_{ij}|^{2}$ ; one can also write $Y_{ij} = \mbox{Tr}(\widetilde \rho_{i} \rho_{j})$.  \cite{Aristov2011a}  

The appropriate representation for the reduced conductance matrix may be constructed by using 
  generators of $U(4)$ Cartan subalgebra, which are three traceless diagonal matrices and one unit matrix. We define 
\begin{equation}
\begin{aligned}
\mu_1 &=1/\sqrt{2}  \mbox{diag}(1,-1,-1,1), \\
\mu_2 &= 1/\sqrt{2} \ \mbox{diag}(1,1,-1,-1),\\
\mu_3 &=1/\sqrt{2}  \mbox{diag}(1,-1,1,-1),\\
\mu_4 &=1/\sqrt{2} \ \mbox{diag}(1,1,1,1).
\end{aligned}
\end{equation}
with the property  $\mbox{Tr}(\mu _{j}\mu _{k})=2\delta _{jk}$,  $j=1,\ldots, 4$. The densities
may be expressed now as $\rho _{j}=1/\sqrt{2}\sum_{k }R_{jk }{%
\mu }_{k }$, where the 4$\times$4 matrix $\mathbf{R}$ is given by
\begin{equation}
\mathbf{R}= \frac 12 \begin{pmatrix}
 1 & 1 & 1 & 1 \\
 -1 & 1 & -1 & 1 \\
  -1 & -1 & 1 & 1 \\
 1 & -1 & -1 & 1 \\
\end{pmatrix}
\end{equation}%
and has the properties $\mathbf{R}^{-1}=\mathbf{R}^{T}$, $\mbox{det }\mathbf{R}=1$. 
The outgoing amplitudes are expressed in  similar form with 
$\mu_{j}$ replaced by $\widetilde{\mu }_{j}=S^{+} \mu _{j} S$. 
It also means \cite{Aristov2011} that we work now with the  combinations of currents and voltages of the form
$I_{i} = \sum_{k} R_{ik} I_{k}^{new}$,  $V_{i} = \sum_{k} R_{ik} V_{k}^{new}$,  or 
\begin{equation}
\begin{aligned}
I_{1}^{new} &= (I_1 - I_2 - I_3 + I_4)/2  \,,\\  V_{1}^{new} &=(V_1 - V_2 - V_3 + V_4)/2  \,,  \\ 
I_{2}^{new} &= (I_1 + I_2 - I_3 - I_4) /2 \,,\\  V_{2}^{new}& =(V_1 + V_2 - V_3 - V_4)/2 \,,  \\ 
I_{3}^{new} &= (I_1 - I_2 + I_3 - I_4)/2 \,,  \\
 V_{3}^{new} &= (V_1 - V_2 + V_3 - V_4)/2  \,,  \\ 
 I_{4}^{new} &=\sum\nolimits_{j} I_{j}/2 \,,  \quad 
 V_{4}^{new}  = \sum\nolimits_{j} V_{j}/2 \,.
\end{aligned}
\label{def:newIV}
\end{equation}
Our notation is slightly different from Ref.\ \onlinecite{Teo2009}, where the Kirchhof's law, $I_{4}^{new}=0$, was explicitly used.   The  meaning of the currents $I_{j}^{new}$, linked to the Fig.\ \ref{fig:spinful},  is as follows: $I_{1}^{new}$ is the charge current moving to the right, $I_{1}^{new}$ - charge current moving down, and $I_{3}^{new}$ is the spin current. 
The resulting reduced conductance matrix in our new basis is determined by $\mathbf{G}=\mathbf{R}^T\,\mathbf{C}\, \mathbf{R} = \frac12 (1- \mathbf{Y} ^{R})$ 
with $Y^R_{ij}=\tfrac{1}{2} \mbox{Tr}(\widetilde{{\mu }}_{i}{\mu} _{j})$ 
  and has a general structure
\begin{equation}
\mathbf{G} =
\begin{pmatrix}
3\times3 & 0 \\
 0 & 0
\end{pmatrix}
\label{defGmat}
\end{equation}
 For our choice of $S$-matrix below in Eq.\  \eqref{genericS},  $\mathbf{G}$ attains even simpler diagonal form.

\subsection{  $S$-matrix for helical edge states}

It was noted in \cite{Teo2009} that in addition to unitarity, $S^{\dagger}S =1$, the time  reversal symmetry leads to the additional condition on the S-matrix. Namely, under the time reversal, ${\cal T}$, one has ${\cal T}\Psi _{in} = E \Psi _{out}$ and  ${\cal T}\Psi _{out} = - E \Psi _{in}$ with $E=diag[1,-1,1,-1]$.   It results in the relation $S = -E S^{T} E$, i.e.\ the matrix $ES$ is antisymmetric. 
Additionally assumed constraints on the form of $S$ refer to the symmetry with respect to interchange of the wires:  $1\leftrightarrow2$ and  $3\leftrightarrow4$, which should not change the observable conductivities. In terms of the matrix $Y_{ij} = |S_{ij}|^{2}$ below it reads as 
$Y = X Y X $ with 
 \begin{equation}
 X  =\begin{pmatrix}0 & 1 & 0 & 0 \\
 1 & 0 & 0 & 0 \\
 0 & 0 & 0 & 1 \\
 0 & 0 & 1 & 0 \end{pmatrix}
 \label{Xmat}
  \end{equation}
These constraints define the $S$- matrix in the form  
 \begin{equation}
  S=  \begin{pmatrix}
0& t & f & r \\
t &  0 & -r^{*} & f^{*}\\
 -f &  -r^{*} &0 & t^{*}\\
 r & -f ^{*} & t^{*} & 0 \\
\end{pmatrix} ,
\label{genericS0}
\end{equation}
with complex- valued $ t , f , r$ and $ |t|^{2} + |f|^{2}+|r|^{2}=1$.  Our analysis below is invariant with respect  to ``rephasing'' operations,   $S \to  U^{1} S U^{2}$ with $U^{1,2}$ arbitrary matrices of the form $U_{kl} = \delta_{kl}e^{i \alpha_{k}}$.
Using such rephasing we can represent the S-matrix for our purposes as  
\begin{equation}
  S=  \begin{pmatrix}
0& t & f & r \\
t &  0 & r e^{iu} &- f  e^{iu}\\
 -f & r e^{iu} &0 &- t e^{iu}\\
 r & f e^{iu} & -t e^{iu}& 0 \\
\end{pmatrix} ,
\label{genericS}
\end{equation}
now with  real-valued $t$, $f$, $r$  subject to the condition $   t^{2} + f^{2}+r^{2}=1$ and arbitrary $u$.  We notice here that the operation $XSX$ corresponds to a change $u \to \pi - u$ in \eqref{genericS}, up to some rephasing. 

We also note that it is always possible to make a partial rephasing of $S$ in \eqref{genericS0} so that $f$ is real-valued, in this case $S$ becomes symplectic matrix from $Sp(2)$ group.  This latter property of S-matrix for the helical edge states is principally different from the previously studied case of X-junction between usual wires, \cite{Aristov2015b} where the mentioned symplectic property is achieved in a very special case, $\alpha_{1}=-\alpha_{2}$, in notations of  Ref.\ \onlinecite{Aristov2015b}.  
  
  We  may parametrize \eqref{genericS} by  two angles as 
 \begin{equation}
t=   \cos   \beta\,,  \quad 
r=\cos\gamma\sin   \beta\, \quad 
f =  \sin\gamma\sin   \beta\,,
\label{Sparam2}
\end{equation} 
which leads to the matrix of  conductances in the form:
\begin{equation}
\begin{aligned}
\mathbf{G} 
 & = 
\tfrac12 (1-\mathbf{Y}^R)=\tfrac12 \mbox{diag}[1-a,1-b,2+a+b] \\
& \equiv \mbox{diag} [G_{R}, G_{D}, G_{S}]
\end{aligned}
\label{redcond}
\end{equation}
where $a=2 \sin ^2 \beta \cos^{2}  \gamma - 1$, $b=  \cos 2 \beta $.  
The region of the allowed values of conductance in the $(a,b)$ plane is given by a triangle defined by vertices 
$(-1,-1)$, $(1,-1)$, $(-1,1)$, as shown in Fig. \ref{fig:4TI2int} below. This statement is initially made for non-interacting fermions, but we also verify below, that the interaction-induced RG flows of the parameters never drive the system beyond this triangle, and the conductances are defined in terms of $a,b$. This means that we have a relation $$ G_{S} = 2-G_{R}-G_{D} .$$
In the absence of spin-flip processes, $f=0$, one has $a=-b$, i.e. the region of allowed conductances is reduced to a segment in $(a,b)$ plane. 

In the presence of interactions, the structure of \eqref{genericS} is unchanged, but the elements vary. 
The main effect in the d.c. limit can be described by the renormalization of the $S$-matrix,  \cite{Aristov2011a} 
which translates to the renormalized quantities $a$, $b$  in Eq.\ \eqref{redcond}. 

\subsection{Renormalization group equations}

The renormalization of the conductances by the interaction is determined by
first calculating the correction terms in each order of perturbation theory.
We are in particular interested in the scale-dependent contributions
proportional to\ $\Lambda =\ln (L/\ell)$, where $L$ and $\ell$ are two lengths,
characterizing the interaction region in the wires. 

The first logarithmic correction for the S-matrix leads to the renormalization group (RG) equation which was obtained in general form in  \cite{Lal2002}. Second-order subleading corrections which might be important in certain cases were analyzed in the language of $S$-matrix in \cite{Aristov2010}. It was shown, however, \cite{Aristov2011a} that RG flow may appear in some of the phases of $S$-matrix, which are irrelevant for the observable conductances. It is thus reasonable to reformulate the RG procedure entirely in terms of conductances, which allows one-to-one correspondence for Y-junction. \cite{Aristov2011a} This unique correspondence between RG equation for S-matrix and RG equation for matrix of conductances $C$ is valid for Y-junction and  may be explicitly lost for X-junction, as shown in \cite{Aristov2015b}. This ambiguity stems from the impossibility to recover the phases of unitary S-matrix, belonging to $U(N)$ group, from the absolute values of its elements, for $N\ge 4$. In our case here we also have this ambiguity in the form of appearance of the phase factor $e^{iu}$ in \eqref{genericS} mixing  left and right isoclinic $S$-matrices. However, in contrast with  \cite{Aristov2015b} we do not have the ambiguity in RG flows for the considered model of interactions $g_{ij}$. 

In lowest order in the interaction the scale dependent contribution to the
conductances is given by \cite{Aristov2012a}
\begin{equation}
C_{jk}=C_{jk}\big\rvert_{\mathbf{g}=0}
+\tfrac{1}{2}\sum\limits_{l,m} \mbox{Tr}\left[\widehat{W}_{jk}\widehat{%
W}_{lm} \right] g_{ml}\Lambda \,,
\label{Rg_1order}
\end{equation}%
where  $C_{jk}\rvert_{\mathbf{g}=0}=\frac12(\delta _{jk}-Y_{jk})$, the $\widehat{W}_{jk}= [\rho_{j}, \widetilde\rho_{k}]$ are a set of
sixteen  $4\times 4$ matrices (products of $\widehat{W}$'s are matrix products), 
$g_{ml}$ is the matrix of interaction constants appearing in \eqref{Ham} and the
trace operation $\mbox{Tr}$ is defined with respect to the 4$\times$4 matrix space
of $\widehat{W}$'s. 

If we  multiply $C_{ij}$ with $\mathbf{R}^{T}$ from the left and $\mathbf{R}$ from the
right then we get the components of $\mathbf{Y}^{R}$ in the form
\begin{equation}
Y_{jk}^{R}=  Y_{jk}^{R} \big\rvert_{\mathbf{g}=0}-\frac{1}{2}\sum\limits_{l,m} 
\mbox{Tr}\left[\widehat{W}_{jk}^{R}\widehat{W}_{lm}^{R}\right]
g_{ml}^{R}\Lambda \,.
\end{equation}
Here  $\widehat{W}_{jk}^{R}=[\mu_{j}, \widetilde\mu_{k}] =  \{\mathbf{R}^{T}\cdot \widehat{\mathbf{W}}%
\cdot \mathbf{R}\}_{jk}$ are again a set of  4$\times$4 matrices, but now we have $\widehat{W}_{jk}^{R}=0$ for $j=4$ or $k=4$, so only nine matrices $\widehat{W}_{jk}^{R}$ are non-zero.
We also defined 
$g_{ml}^{R}=\{\mathbf{R}^{T}\cdot \mathbf{g} \cdot \mathbf{R}%
\}_{ml}$. The nine nonzero matrices $\widehat{W}_{jk}^{R}$ are   evaluated with the aid of computer algebra.
Differentiating these results with respect to $\Lambda $ (and then putting $\Lambda =0$) we find the RG equations
\begin{equation}
\frac{d}{d\Lambda }Y_{jk}^{R}=-\frac{1}{2}\sum \limits_{l,m}
\mbox{Tr}\left[\widehat{W}_{jk}^{R}\widehat{W}_{lm}^{R}\right] g_{ml}^{R} \,.
\label{eq:RGgeneral}
\end{equation}%

In higher orders of perturbation theory in  $g_{ml}$ we find subleading contributions of two types. \cite{Aristov2011a} One type, appearing first in the third order, provides a three-loop contribution to RG equations and does not influence the scaling exponents around the RG fixed points (FP). Second type of 
contributions is more important, it is given by the ladder sequence of diagrams, and defines the scaling exponents around FPs. Due to peculiarities of 1D model with linear dispersion, each diagram in this ladder sequence is formally a one loop contribution, providing subleading linear-in-$\Lambda$ corrections. After the summation of the ladder series \cite{Aristov2012,Aristov2013} (for diagonal $g_{ij} =g_{i} \delta_{ij}$) one obtains the  renormalized interaction matrix $\bar{\mathbf{g}}$ replacing the bare
interaction matrix $\mathbf{g}$ \ in Eq.(\ref{Rg_1order}). The components of $\bar{\mathbf{g}}$ are obtained from the following matrix equation
\begin{equation}
\bar{\mathbf{g}}=2(\mathbf{Q}-\mathbf{Y})^{-1}\,.
\label{eq:gladder}
\end{equation}%
The matrix $\mathbf{Q}$ characterizes the interaction strength and
depends on the Luttinger parameters $K_{j}=[(1-g_{j})/(1+g_{j})]^{1/2}$ as
\begin{equation}\label{eq:qdef}
Q_{jk}=q_{j}\delta _{jk},\quad q_{j}=(1+K_{j})/(1-K_{j})\,.
\end{equation}

\section{Tunneling between  edge states \label{sec:flows}}

 \subsection{Symmetric corner junction  \label{sub:teo}}

For quantum spin Hall insulator the channels correspond to four helical edge states, with possible tunneling contact between them. The  interaction between the different edge states is absent.  In other words, the wires 1 and 4, 2 and 3 do not interact, and we have the simple interaction matrix:
\begin{equation}
 \mathbf{g} =  g\,  \mathbf{1} \, .
 \label{diag-g}
\end{equation}

The first-order RG equations for S-matrix of the form \eqref{genericS} are trivial  :
\begin{equation}
\frac{d \mathbf{Y} ^{R} }{d\Lambda}=0 \,,
\end{equation}
which is the result of the diagonal form of the interaction \eqref{diag-g} and the absence of backscattering.~\cite{Lal2002}

To advance further we take into account  the higher orders of interaction as explained above,  i.e.\  we replace $\mathbf{g}$ by $\bar{\mathbf{g}}$~\eqref{eq:gladder}. In such a way we obtain non-trivial RG equations:
\begin{equation}
\begin{aligned}
\frac{d a}{d\Lambda}&=\left(\frac{b+1}{b (K-1)+K+1}+\frac{a-1}{a (K-1)+K+1}
\right.\\
& \left. +\frac{a+b}{(a +b)(K-1)-2} \right)(a+1) (K-1), \\
\frac{d b}{d\Lambda}& =  \frac{d a}{d\Lambda} \Big | _{a \leftrightarrow b} \,.
\end{aligned}
\label{RG:fullSym}
\end{equation}
Expanding these RG equations to the lowest non-vanishing order  $\sim g^{2}$  we 
recover the Eq.\ (3.40) in Ref.\ [\onlinecite{Teo2009}].  The second-order diagrams in Fig.\  9 there correspond exactly to the truncation of the series leading to our Eq.\ \eqref{RG:fullSym}. 
It is worth noting that the phase $u$ is absent in \eqref{RG:fullSym}, it is present in $\widehat{W}_{jk}$ in Eq.\ \eqref{Rg_1order} but disappears in the quantity $\mbox{Tr}\left[\widehat{W}_{jk}\widehat{W}_{lm}\right]$.

The RG equations \eqref{RG:fullSym} reveal seven fixed points: three of them are stable for any value of Luttinger parameter, the others are unstable (see Table~\ref{tab:TK}). This result is in agreement with second order calculation in  \cite{Teo2009} and the correspondence between our notation and  Ref.\ \cite{Teo2009}  is  $\mathcal{R}=(1+a)/2$, $ \mathcal{T}=(1+b)/2$, $\mathcal{F}=-(a+b)/2$.
 
 \begin{table}
\caption{\label{tab:TK}  Fixed points for symmetric corner junction. The positions of FPs are given by the coordinates $(a, b)$ in the plane of conductances, the stability of the FPs is also shown. }
\begin{ruledtabular}
\begin{tabular}{c|rrrrrrr}
$a$& -1&-1&-1& -1/3&0&0&1  \\
$b$&  -1&0&1& -1/3&0&-1&-1  \\ \hline
stability &  s&u&s& u&u&u&s \\  
\end{tabular}
\end{ruledtabular}
\end{table}

Each of these FPs is characterized by generally two scaling indices in the plane $(a,b)$, depending on the direction of RG flow with respect to the given FP. 
We found that  three stable FPs (vertices of the triangle) have the same exponent in both directions, equal to  
\begin{equation}
\alpha_{1}= 2-1/K-K <0, 
\label{eq:scal0}
\end{equation}
which corresponds to the weak tunneling process between the TLL wires and agrees with the result in \cite{Teo2009}. 

One unstable FP at the triangle's center is characterized by rather unusual exponents, equal to  $3+27/(2+K)^2-18/(2+K)> 0$,  in both directions.  Three other FPs  residing at the middle of the triangle's edges show the scaling exponents  $2+8/(1+K)^2-8/(1+K) > 0$ and  $3-K-4/(1+K) <0 $ in the direction along and perpendicular to the edge. The latter FPs are of the  saddle point type.  

\begin{figure}[t]
\includegraphics[width=0.85\columnwidth]{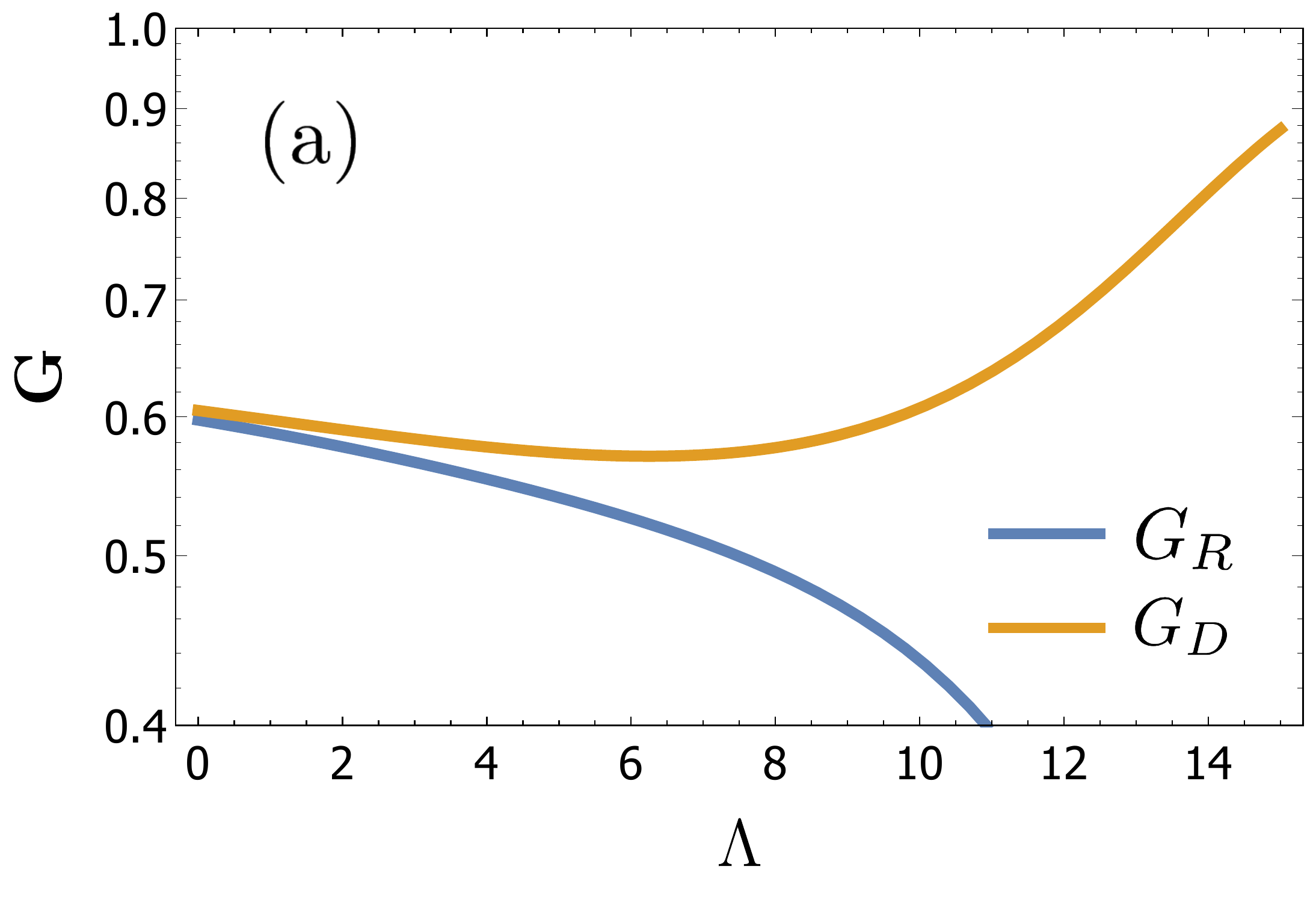}  
\includegraphics[width=0.85\columnwidth]{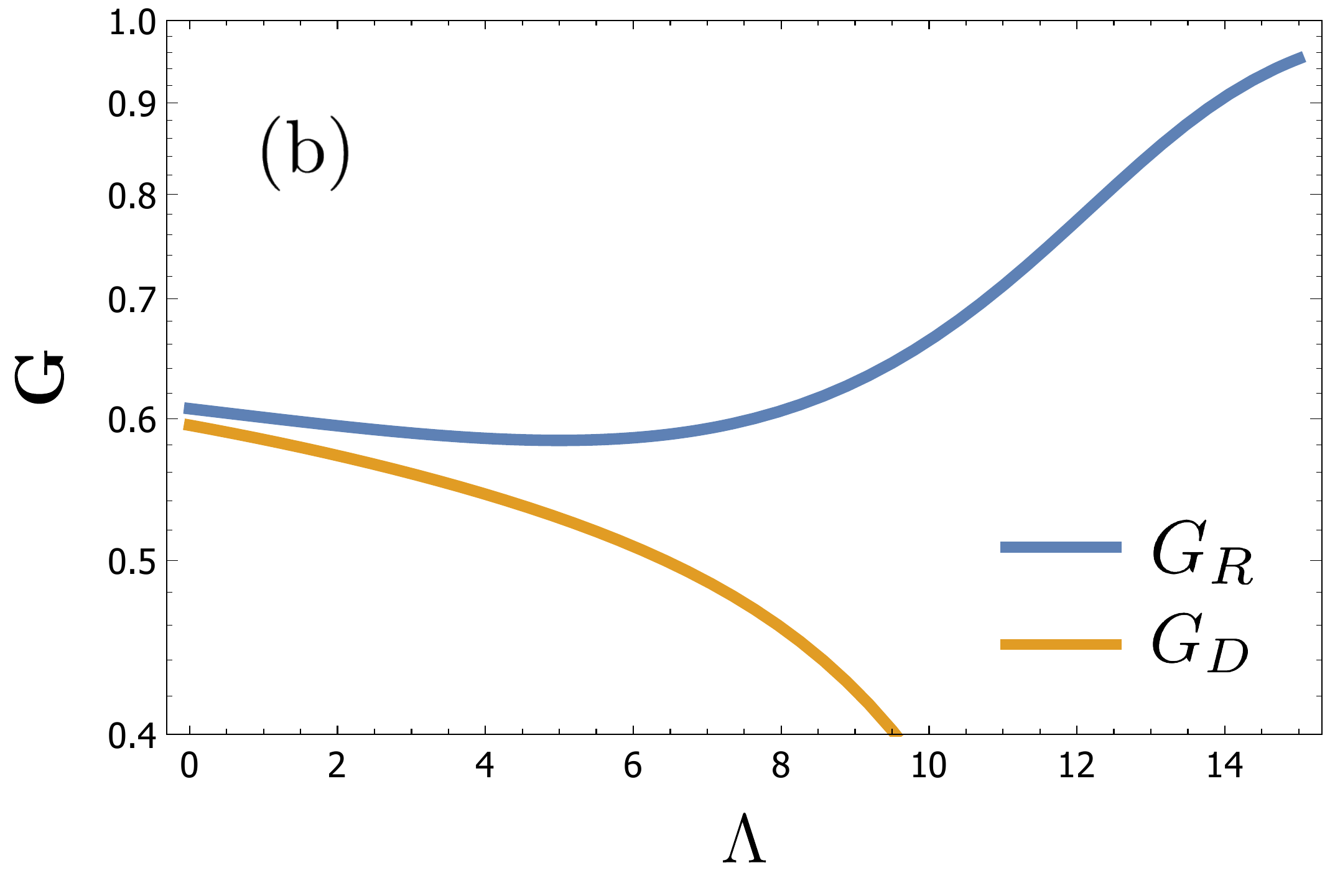}    
\caption{  
Full scaling curves for conductances $G_R=I_1^{new}/V_1^{new}$ and $G_D=I_2^{new}/V_2^{new}$, defined in Eq.\ \eqref{def:newIV}, and the  Luttinger parameter $K=0.4$. Slight difference in the initial conductances results in the opposite qualitative behavior.    The panel (a) and (b) show RG flows tending to the FP of the perfect transmission in the vertical and horizontal direction, respectively, in terms of Fig.\ \ref{fig:spinful}. 
The non-monotonic behavior stems from the RG flow passing near the FP of saddle point type at $a=b=0$ in Fig.\   \ref{fig:4TI2int}. 
}
\label{fig:FullCurve}
\end{figure}

It was shown in \cite{Aristov2010} that the full scaling curves for conductances in case of Y-junction may reveal non-monotonic behavior with the scaling parameter $\Lambda$.  We note here that such behavior should generally happen if the RG trajectories pass near the saddle point FP. To demonstrate this we plot  the full scaling curves for the present case of X-junction in Fig.\ \ref{fig:FullCurve}, choosing reasonable values for the Luttinger parameter and the appropriate values of bare conductances.  

\subsection{Asymmetric corner junction. \label{sub:teotwointeraction}}

In this subsection we consider more general setup,  allowing different strength of interaction in the upper (1-2) and lower (3-4) edge states in Fig.\  \ref{fig:spinful}b.  It is interesting to observe that when the edge states are non-equal,  one could expect more complicated expression for the S-matrix instead of Eq.\  \eqref{genericS}. However the above general arguments leading to Eq.\  \eqref{genericS} involve only  a few discrete symmetries  which are not broken by the non-equivalence of the upper and lower edge states.  This means that the system remains at a fixed surface in space of conductances, which is 
protected with respect to asymmetric perturbations of the S-matrix. In other words, only consistent changes in the transmission coefficients within the wires 12 and 34 are allowed, so that $|S_{12}| = |S_{34}| = t$. This may be contrasted with interaction-induced asymmetric perturbations in case of  Y-junction, which can drive the system from the symmetrical point, as discussed in  Sec. VI of Ref.\ \onlinecite{Aristov2013}.

\begin{figure}[t]
\includegraphics[width=0.85\columnwidth]{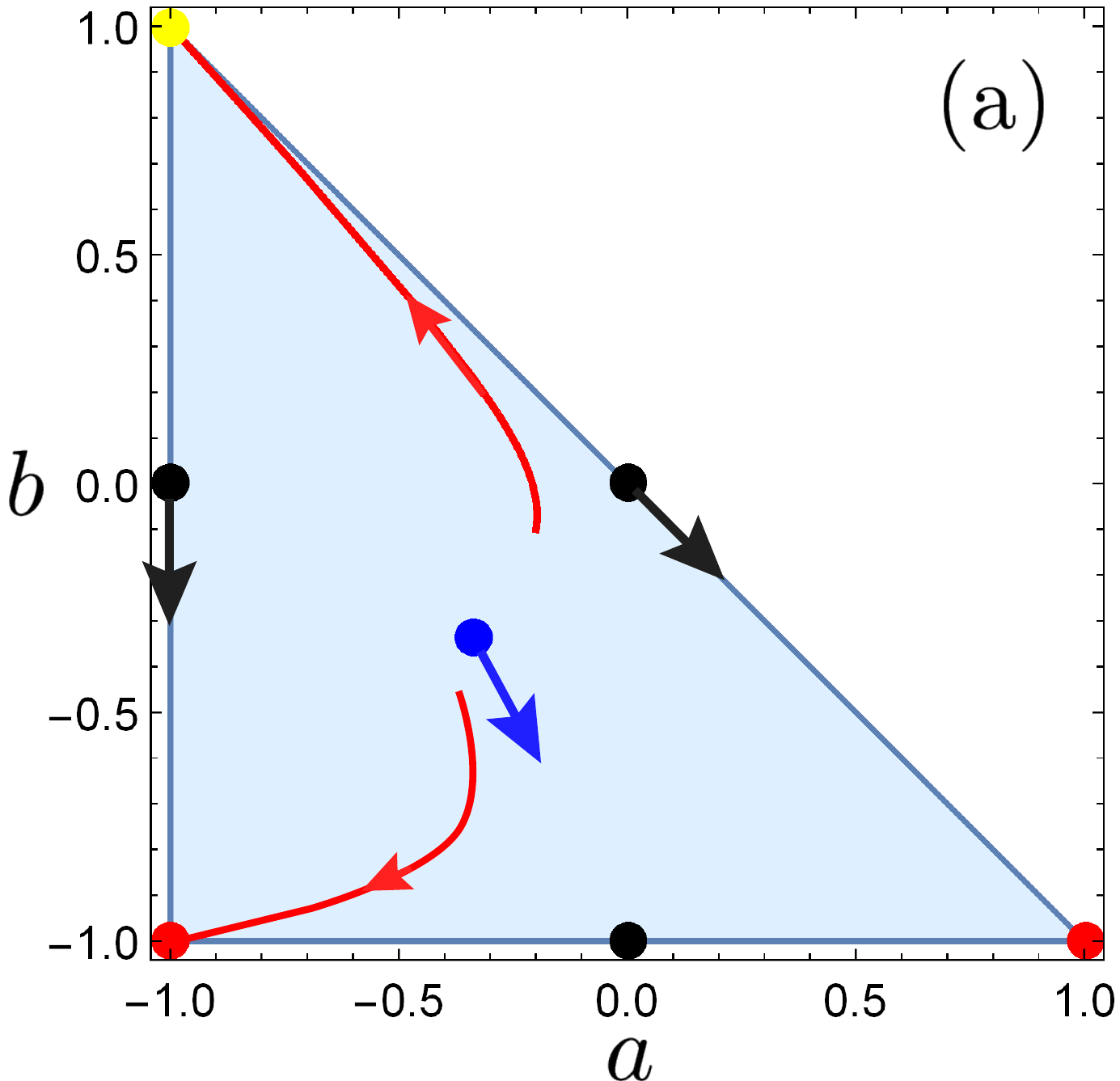}   
\includegraphics[width=0.85\columnwidth]{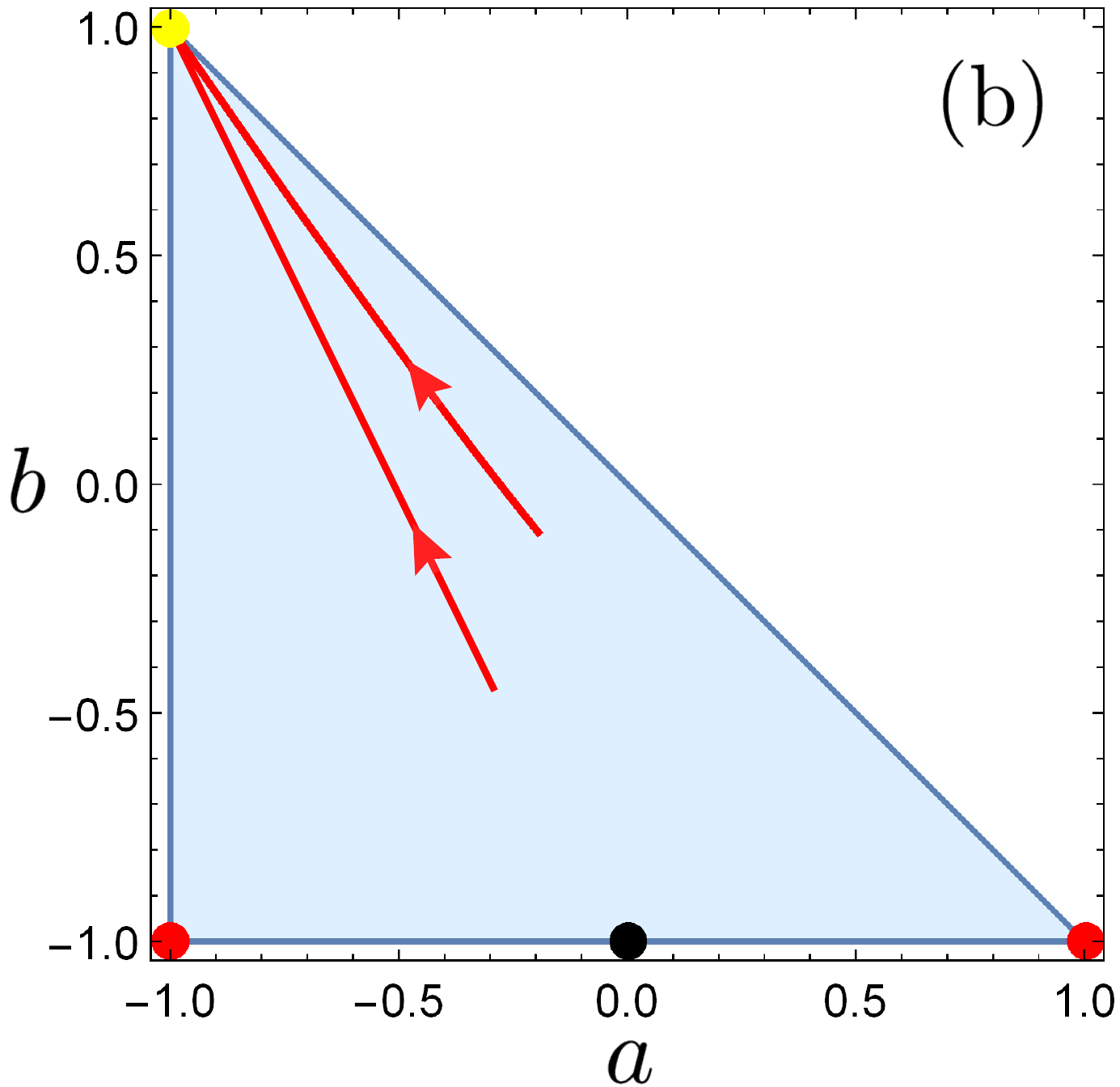} 
\caption{RG flows (red lines, the direction of flow shown by arrows) and fixed points are  shown for different  values of interaction in the edges states. The panel   (a) corresponds to Luttinger parameters  $K_1=0.3$, $K_2=0.5$ and panel (b) is for $K_1=1.2$, $ K_2=0.5$. The disappearance of three non-universal unstable FPs is visible in panel (b), which is accompanied by the change of the character of the lower red FPs. }
\label{fig:4TI2int}
\end{figure}

We take the matrix of the dimensionless 
interaction constants in the form :
\begin{equation}
 \mathbf{g} =  \text{diag}[g_1,g_1,g_2,g_2]  \, ,
 \label{diag-g_2}
\end{equation}
and use the previous $S$-matrix \eqref{genericS}. The result of the ladder summation, Eqs.\ \eqref{eq:RGgeneral}, \eqref{eq:gladder}  leads to rather complicated RG equations. In order to illustrate the qualitative picture, we first expand the coupled RG equations to  the second order of interaction  and write 
\begin{equation}
\begin{aligned}
\frac{d a}{d\Lambda}&=- \frac18 (1 + a) ((1 + b)^2 g_1^2 \\
&+ 2 (-1 + 2 a^2 + 2 a b + b^2) g_1 g_2 + (1 + b)^2 g_2^2), \\
\frac{d b}{d\Lambda}&=- \frac18 (1 + b) ((-1 + b^2) g_1^2 \\
&+ 2 (1 + 2 a (1 + a) + 2 a b + b^2) g_1 g_2 \\
&+ (-1 + b^2) g_2^2)\,.
\end{aligned}
\label{eq:RGasym}
\end{equation}
We observe  that these equations  formally have seven FPs, but these FPs not always lie in the physical region. Three FPs with $ -1 < b < 1$ are now non-universal  and their position depends on the ratio of the interactions $g_{1}/g_{2}$ as shown in and Table \ref{tab:TK2}.  We also show the tendency of these intermediate FPs to change their position in Fig.~\ref{fig:4TI2int}a.  The two previously stable FPs at $a = \pm 1$ , $b =-1$ keep their position but their stability now also depends on $g_{1}/g_{2}$.  As can be seen from Table \ref{tab:TK2}, if $g_{1}/g_{2} > 0$ and in particular $g_{1} = g_{2}$ then all three non-universal points lie either within or on the border of the triangle of physically allowed conductances. If $g_{1}/g_{2} < 0$ then non-universal points disappear in the physical region and this is accompanied by the change in the character of some of the remaining FPs.

\begin{table}
\caption{\label{tab:TK2} The position of non-universal fixed points for the case of non-equal interaction in the wires. }
\begin{ruledtabular}
\begin{tabular}{c|ccc}
     & edge & edge &  median  \\ \hline
$a$& $-1$&$\frac{(g_1-g_2)^2}{(g_1+g_2)^2}$&$ -\frac 13+ \frac13 
\frac{(g_1-g_2)^2}{g_1^2+g_1 g_2+g_2^2} $\\
$b$& $  -\frac{(g_1-g_2)^2}{(g_1+g_2)^2}$&$-a$&$-1-2a  $
\end{tabular}
\end{ruledtabular}
\end{table}

\begin{figure}[t]
\includegraphics[width=0.9\columnwidth]{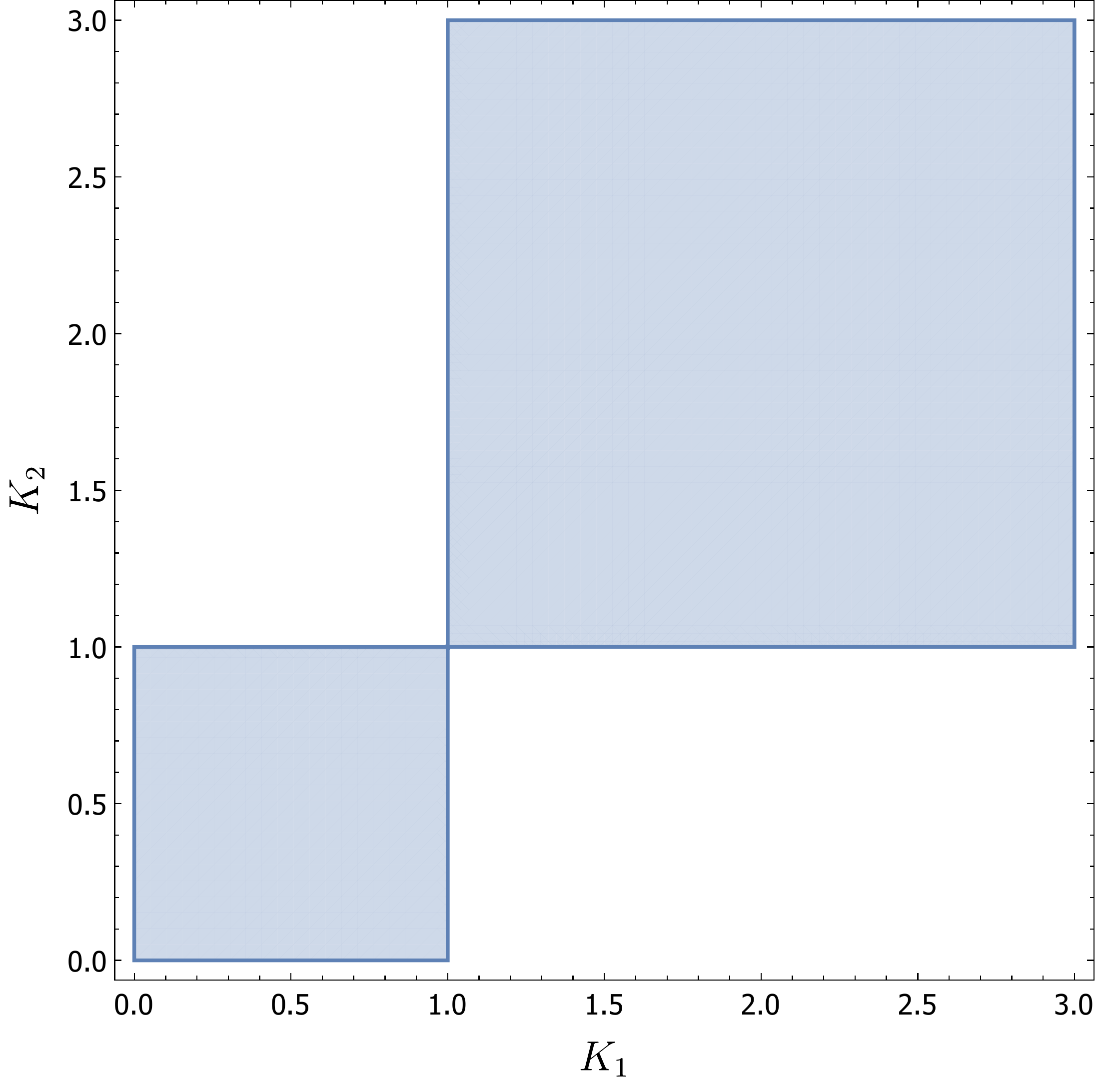} \caption{Phase portrait for conductances in a setup with  different  values of interaction in the edges states, characterized by Luttinger parameters, $K_{1,2}$. Depending on the sign of $(1-K_{1})(1-K_{2})$ one has either one or three stable fixed points, as described in text. 
}
\label{fig:TI2PhasePortrait}
\end{figure}

This qualitative picture obtained in the second order of perturbation is confirmed by the analysis 
of the full form of the RG equations.
Summarizing here, 
we have two different regions in the plane of Luttinger parameters $K_{1}, K_{2}$ (see Fig.~\ref{fig:TI2PhasePortrait}). The blue region is characterized by the  existence of  seven  FPs, and three of these points are stable, whereas other three FPs are non-universal in their position. This region remains in qualitative correspondence with the case of equal interactions discussed in the previous subsection.  The new region marked white is characterized by  the existence of four FPs, one of them is stable, two (previously stable) FPs are unstable, and the fourth point remains of the saddle point character. The only stable point in this white region corresponds to fully disconnected upper and lower wires, which are perfectly transmitting within themselves. 
On the lines dividing  white and blue regions in  Fig.\ \ref{fig:TI2PhasePortrait} 
three non-universal FPs coincide with three lowest FPs in Fig.\ \ref{fig:4TI2int}, i.e.\  $a=-1,0,1$, $b=-1$. 
The largest value of $b$ for non-universal FPs is achieved when the interactions are equal, $g_{1}=g_{2}$. 

The yellow FP in  Fig.\ \ref{fig:4TI2int}, $a=-b=-1$,  is always stable and corresponds to disconnected lower and upper wires; it is  CC-point
in notation of the work \cite{Teo2009}.    The right red FP in  Fig.\ \ref{fig:4TI2int}, $a=-b=1$, corresponds to the perfect transmission in the vertical direction of Fig.  \ref{fig:spinful} ; it is II -point in in notation of  \cite{Teo2009}.  The left  red FP in  Fig.\ \ref{fig:4TI2int}, $a=b=-1$, corresponds to spin current conductance $G_{S}=0$ and was called a perfect spin flip transmission point in \cite{Teo2009}.   

Summarizing here,  we list the scaling exponents of the conductances near the corresponding FPs of Eq.\ \eqref{eq:RGasym}.  
Near the  points  $(a,b)=(\pm1,-1) $ we have the same exponent equal to $  {-2 (K_2-1)( K_1-1)}/{(K_1+K_2)}$;   near $(a,b)=(-1,1)$  we have the exponent $-\frac{1}{2} \left( ( K_1-1)^{2}/K_{1} + (K_2-1)^{2}/ {K_2} \right)$. Near the unstable (universal) FP, $(a,b)=(0,-1)$,  we find two different exponents, $ {2 (K_1-1) (K_2-1)}/{((K_1+1) (K_2+1))} $ and  
 $ -\frac{(K_1-1) (K_2-1) (2 K_2 K_1+K_1+K_2)}{(K_1+1) (K_2+1) (K_1+K_2)}$ along $a$ and $b$, respectively.



\section{Tunneling to helical edge state from spinful wire   \label{sub:probe}}

As was argued above, the form of 
 $S$-matrix \eqref{genericS}  does not imply the equivalence between the upper and lower wires. In particular,  we may view  the lower wire $3-4$ as the tip of the spinful wire, whereas the channels 1 and 2 remain associated with the helical edge states. In the absence of the tunneling between this tip and the edge state we have $f=r=0$, $t=1$ which in the usual sense corresponds to a perfect reflection of the electrons at the end of the spinful tip. 
 
 The main difference of this setup is a new form of  interaction  matrix:
\begin{equation}
 \mathbf{g} =\begin{pmatrix}
  g_1, &0, & 0, & 0   \\
  0, &g_1, & 0, & 0   \\  
  0, &0, & g_2/2, &  g_2/2   \\  
  0, &0, &  g_2/2, &  g_2/2  
\end{pmatrix} ,
\end{equation}
which  describes the charge-charge interaction in the spinful wire.  The non-diagonal  form of $ \mathbf{g}$ requires certain adjustment of our analysis.  The first-order RG equation \eqref{eq:RGgeneral} is unchanged and the result of summation of higher order terms \eqref{eq:gladder}  should be revised. 
The appropriate way of doing it can be found in the idea of the spin-charge separation in the spinful wire tip and the details of our derivation of \eqref{eq:gladder} in \cite{Aristov2012,Aristov2013}. We notice that  dressing of interaction $g_{j} \to K_{j}$ happens in the bulk of the wire away from the contact, and does not involve $S$-matrix. The whole procedure of such dressing for non-diagonal $\mathbf{g}$ can be then performed by first diagonalizing the interaction by appropriate orthogonal transformation, thus passing to a description in terms of charge and spin density in the spinful wire. Second step involves the dressing of the bulk interaction effects, by summing the ladder diagrams, it is encoded in matrix quantity $\mathbf{C}$  in \cite{Aristov2012,Aristov2013}. The third step consists of the inverse orthogonal transformation and properly consideration of the contact, described by the above matrix $\mathbf{Y}$. 
Introducing the orthogonal matrix $V = (E + X) / \sqrt{2} = V^{T}=V^{-1}$, with $E$, $X$ defined  in Eq.\eqref{Xmat} and before  it, this sequence of steps is shown schematically as 

\begin{equation}
\begin{aligned}
\mathbf{g} &  \to V \mathbf{g} V  \to 2  \bar{\mathbf{Q}}^{-1} 
\\ & \to  
2\bar{\mathbf{Q}}^{-1}(1+ \mathbf{Y}\bar{\mathbf{Q}}^{-1}+\mathbf{Y}\bar{\mathbf{Q}}^{-1}\mathbf{Y}\bar{\mathbf{Q}}^{-1}+ \ldots ) 
\\ & 
\to   2V\bar{\mathbf{Q}}^{-1} V (1-  \mathbf{Y} V\bar{\mathbf{Q}}^{-1}V )^{-1} 
\end{aligned}
\label{dressing}
\end{equation}
now with $\bar{\mathbf{Q}}^{-1} = diag[q_{1}^{-1},q_{1}^{-1},q_{2}^{-1},0]$ 
and $q_{j}$ defined in \eqref{eq:qdef}. 
Overall, instead of \eqref{eq:gladder}  we write 
\begin{equation}
\bar{\mathbf{g}}=2(V \bar{\mathbf{Q}} V -\mathbf{ Y})^{-1}\,.
\label{eq:gladder2}
\end{equation}
with the singularity in $\bar{\mathbf{Q}}$ is resolved according to the last line in \eqref{dressing}. 
  
Using the reduced conductances, Eq.\ \eqref{redcond} we find the RG equations in the form:
\begin{eqnarray}
\frac{d a}{d\Lambda}&=& (a+1) F(b)  \,,  \nonumber \\
\frac{d b}{d\Lambda}&= & (b-1) F(b)  \,,   
\label{eq:RGspinful}   \\
F(b) & = & \frac{2 (b+1) \left(b (q_1+q_2-2)+q_1^2-q_1+q_2-1\right)}{(b+2 q_1+1) (b (q_1+q_2-2)-2 q_1 q_2+q_1+q_2)} \,.   \nonumber
\end{eqnarray}
It follows then 
that the set  \eqref{eq:RGspinful} is in fact reduced to second equation, and 
\begin{equation}
 \frac{d }{d\Lambda} \frac{a+1}{b-1} = 
 \frac{d }{d\Lambda} \frac{1-G_{R}}{G_{D} } 
 =0 \,,
\end{equation} 
It means that the RG flows in the plane $(a,b)$ lie on straight lines passing through the point  $(-1,1)$ and the ratio, $\frac{a+1}{b-1}$, defined for bare quantities, see Fig.\ \ref{fig:probeab} below. In terms of conductances  
\eqref{redcond} we see that the ratio $(1-G_{R})/G_{D}$ is unchanged under renormalization, i.e. the ``up-to-down''  conductance remains proportional to the value of ``left-to-right''  conductance complementary to conductance quantum. 

\begin{figure} [t]
\includegraphics[width=0.85\columnwidth]{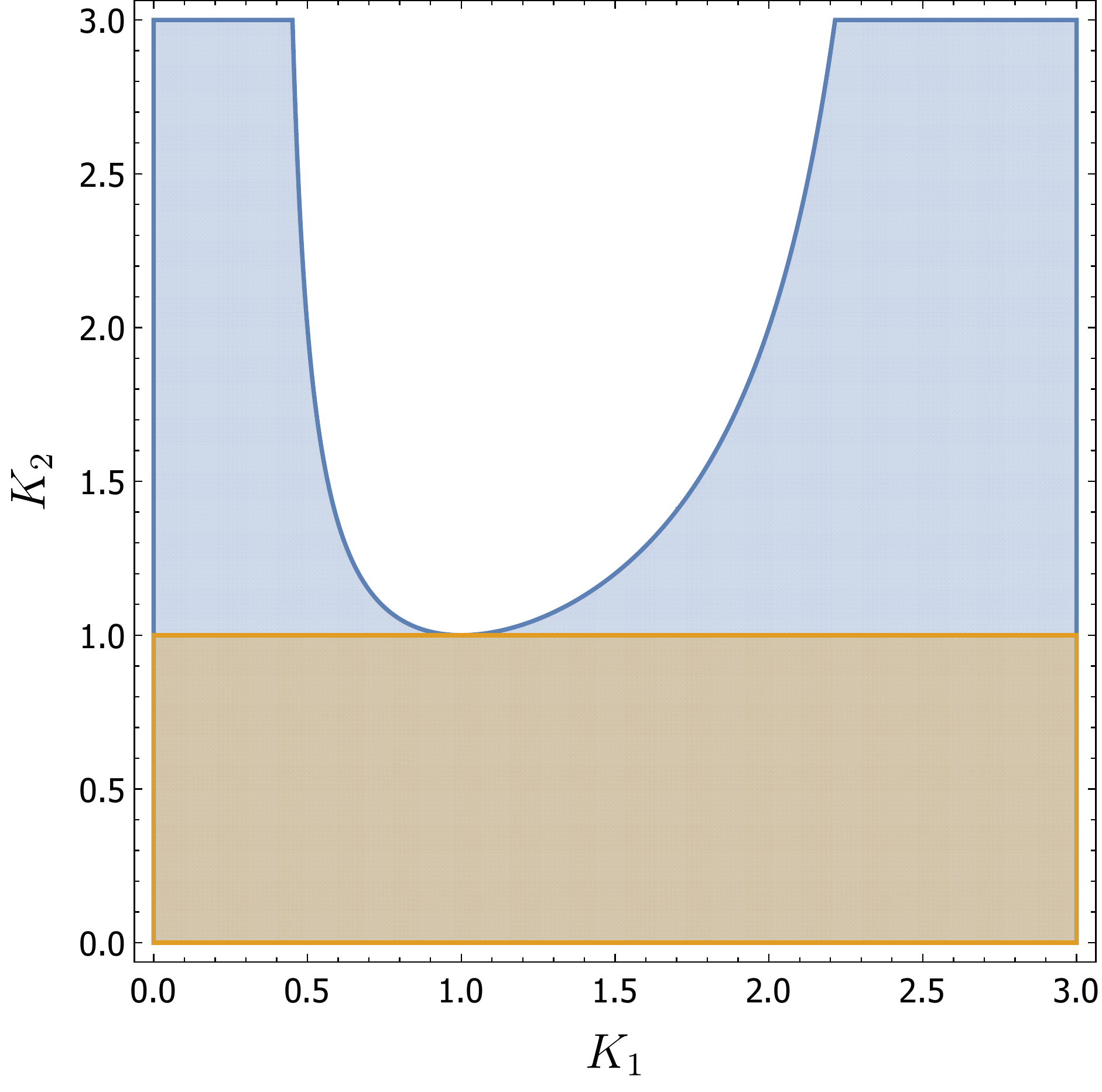} \caption{RG phase portrait for tunneling between the spinful wire and helical edge state.  The brown region at $K_{2} < 1$  is the stability region for the FP corresponding to the absence of tunneling. In the blue region the situation is characterized by one stable FP and a stable fixed line, as shown in Fig. \ref{fig:probeab}b. In the white region only the fixed line $b=-1$ is stable. }
\label{fig:probestab}
\end{figure}

The dependence of $F(b) $ on the interaction is not transparent and we  show a few first terms in powers of $g_{j}$.   
The expansion of \eqref{eq:RGspinful} begins with the second order $g_{1}$, and with the  first order in $g_{2}$. Keeping these  lowest order terms, we have 
\begin{equation}
\begin{aligned}
\frac{d a}{d\Lambda}&=-\frac{1}{4}  (a+1) (b+1) \left( g_2 +
\tfrac{1}{2}   g_1^2  (b+1) \right) \\
\frac{d b}{d\Lambda}&=\frac{1}{4}  (1-b^2) \left( g_2 +
\tfrac{1}{2}   g_1^2  (b+1) \right)\,.
\end{aligned}
\end{equation} 
These equations are rather unusual, because of the existence of one FP and one or two fixed lines. 
The fixed point is defined by $a=-1$, $b=1$, corresponds to the absence of tunneling to the edge state and  is stable everywhere except for the region 
$g_{2}< - g_{1}^{2}$.  
The analysis of the full expression shows that this FP becomes unstable at $K_2> K_1/(3 K_1 - K_1^2-1)$, this is depicted by white region  in  Fig.~\ref{fig:probestab}. The poles of the latter expression determine a finite range of interaction $K_{1}$ in the edge state when the stability of the FP may be lost, namely $\frac{1}{2} \left(3-\sqrt{5}\right)<K_1<\frac{1}{2} \left(\sqrt{5}+3\right)$.

In addition, we have two fixed lines, one at $b=-1$, which is unstable for $g_{2}>0$ (i.e. at $K_{2}<1$, shown as brown region in Fig.  \ref{fig:probestab}) and is stable otherwise. 
The second fixed line at $b=b_{0}$ is determined by the condition $F(b_{0})=0$ which gives  $b_{0}=-(q_{1}^2- q_{1} + q_{2}  -1)/( q_{1} + q_{2}-2 )$, revealing its non-universal character. It is unstable and is indicated as the black line in Fig.~\ref{fig:probeab}b.  The domain of existence of this fixed line correspond to a blue region in Fig. \ref{fig:probestab}. 

We find the scaling exponent around the FP $(a,b)=(-1,1)$  given by 
\begin{equation}
\alpha_{FP}  = \tfrac{1}{2} \left(-K_1-\tfrac{1}{K_1}-\tfrac{1}{K_2}+3\right) \,, 
 \label{eq:scal1}
\end{equation}
and the scaling exponent near the fixed line at $b=-1$ is 
\begin{equation} 
\alpha_{FL}  = -2\frac{K_1^2 (K_2-1)}{(K_1+1) (K_1+K_2)} \,.
 \label{eq:scal2}
\end{equation}
The border between the blue and white region   in Fig. \ref{fig:probestab}  corresponds to a condition $b_{0} = 1$ which translates to $\alpha_{FP} =0$. Similarly, the border between the brown and blue region in Fig. \ref{fig:probestab} is given by the condition $b_{0} = -1$ and $\alpha_{FL} =0$, i.e. $K_2=1$. 

Two examples of RG flows are shown in Fig. \ref{fig:probeab}, where the upper panel corresponds to the  brown region in Fig. \ref{fig:probestab} and the lower panel, Fig. \ref{fig:probeab}b - to the blue region in Fig. \ref{fig:probestab}.    If we fix $K_{1}$ and increase $K_{2}$, then at smallest $K_{2}$ we see the qualitative picture  of  RG flows shown in  Fig. \ref{fig:probeab}a. Then at $K_{2}=1$ the second fixed line appears in the triangle of physical  conductances in  $(a, b)$ plane. Upon further increase of $K_{2}$ this line moves upwards
where the RG flows follow the pattern of Fig. \ref{fig:probeab}b. Finally, the non-universal fixed line crosses the FP with  $a=-1$, $b=1$, and disappears; this corresponds to the white region in Fig. \ref{fig:probestab}, where only the line $b=-1$ is  stable.

\begin{figure}[t]
\includegraphics[width=0.85\columnwidth]{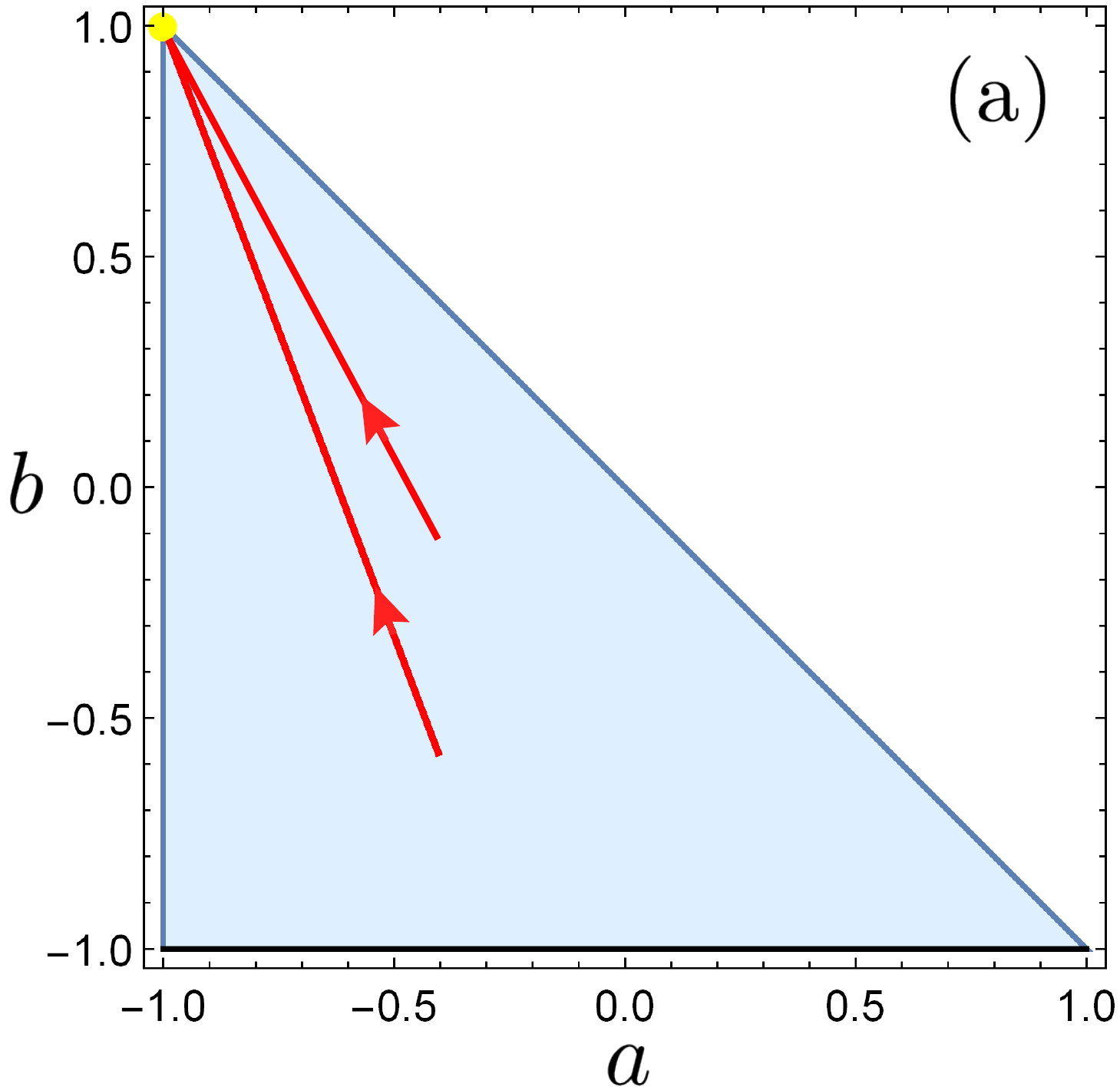}  
\includegraphics[width=0.85\columnwidth]{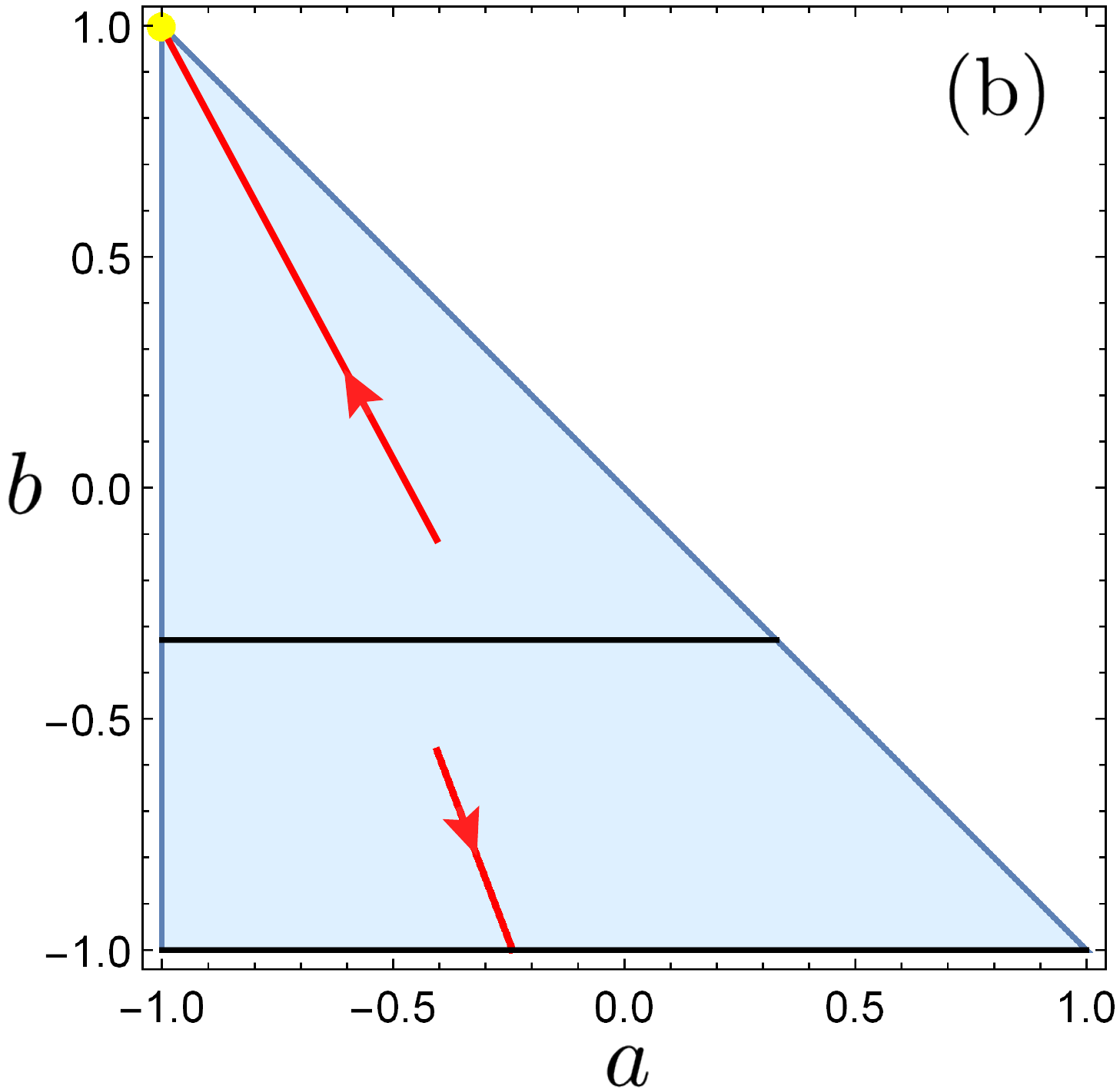} 
\caption{Two examples of RG flows are shown. Typical behavior in the brown region of Fig. \ref{fig:probestab} is shown in panel (a) for $K_1=1.55$, $K_2=0.51$. The situation in blue region of Fig. \ref{fig:probestab} is qualitatively reproduced in panel (b) at $K_1=1.38$, $ K_2=1.03$.}
\label{fig:probeab}
\end{figure}

We compare our findings \eqref{eq:scal1} with the bosonization approach, particularly with  Ref.\  \cite{Teo2009}, where the   
the  scaling exponent \eqref{eq:scal0} of the conductance in the weak backscattering limit was found. 
The exponent in this case can be regarded as  twice the scaling dimension of the tunneling operator at the edge state,  $(K-1)^{2}/2K$.  In the considered setup one TI is substituted by the semi-infinite spinful wire. The contact between a semi-infinite quantum wire and a 3D metal  was analyzed in \cite{Fabrizio1995}, with the  conductance scaling exponent  found as $\frac12 (K_\rho^{-1}+K_\sigma^{-1})-1$. For spin isotropic interaction in our case we should put $K_\sigma=1$ and  $K_\rho=K_2$. Combining both cases we obtain the overall scaling exponent in the form $(K_1+1/K_1)/2+(1+1/K_2)/2-2$ which expression exactly corresponds to Eq. \eqref{eq:scal1}. 

Summarizing here, we obtained the phase portrait for the case of the tunneling from the spinful wire tip to the helical edge state, we find the scaling exponents which are in agreement with bosonization approach when available. 
We note that the appearance of the RG fixed lines in the plane of available conductances  is rather unusual phenomenon, discussed previously in the case of  chiral Y-junction for certain relations between the interaction parameters. \cite{Aristov2012a}

\section{Conclusions \label{sec:conclu}} 

In this paper we consider four-terminal junctions, involving helical edge states of topological insulators.  The time reversal symmetry arguments for such junctions define the general structure of single-particle S-matrix, describing the tunneling processes between the edges. In presence of interactions the conductances characterizing the junction are renormalized, which effect is described in the proposed formalism by a set of non-perturbative RG equations. In order to validate our approach we first  analyze the setup with two helical edge states and show that our results agree with the bosonization studies, when available.  The fixed point structure, scaling exponents and the overall phase portrait of this setup is obtained. 

We further observe that the same symmetry arguments can be applied to the S-matrix for the tunneling from the tip of spinful wire to the helical edge state. This important physical setup was not previously considered. The difference of this setup from the contact between two helical edge states is in the form of interaction, which requires certain modification of our formalism.  The phase portrait is now more involved and, depending on the interaction,  we find the possibility of RG fixed points and fixed lines of conductances. We show that the calculated scaling exponents coincide with those expected from bosonization.  We predict that in the discussed setup the scaling is defined by one equation, and certain proportionality relations between three different conductances are obeyed during renormalization. This prediction may hopefully be checked in future experiments. 
 
The advantage of the employed S-matrix approach is the possibility to obtain both the full scaling curves for the conductances and exact scaling exponents at the fixed points. We  demonstrate that if the RG flow drives the system near by unstable fixed points then non-monotonic scaling behavior of the conductances is observed.   
 
\acknowledgements

We are grateful to P. W\"olfle for numerous useful discussions. This work was supported by the Russian Scientific Foundation grant (project 14-22-00281) and by RFBR grant No 15-52-06009.


\end{document}